# The AI Codebase Maturity Model: From Assisted Coding to Fully Autonomous Systems

## A Practitioner Experience Report from KubeStellar Console and Hive

**Andy Anderson, Ph.D. (@clubanderson)** IBM Research — Hybrid Cloud and AI Platform

*Dr. Anderson holds a Ph.D. in Technology Adoption and has been with IBM for 9 years, the last 4 with IBM Research. He has served as chief maintainer of the KubeStellar organization for 4 years, stewarding it through 3 years as a CNCF Sandbox project.*

**Transparency note:** This paper was written with AI assistance. The author was interviewed by Claude Code (Anthropic's Claude Opus), and the content was organized with AI support. The 4-month experience described herein for Levels 1–5, and the initial deployment of Hive for Level 6, are entirely the author's — every architectural decision, every failure, every insight came from hands-on daily work with AI coding agents. Using AI to write about AI-driven development felt appropriate, and honestly, it would have been strange not to.

---

## Abstract

AI coding tools are widely adopted, but most teams plateau at "prompt and review" without a framework for systematic progression. This paper presents the AI Codebase Maturity Model (ACMM), a **6-level framework** describing how codebases evolve from basic AI-assisted coding to fully autonomous systems. Inspired by CMMI, each level is defined by its feedback loop topology — the specific mechanisms that must exist before the next level becomes possible. I validate the model through a 4-month experience report maintaining KubeStellar Console, a CNCF Kubernetes dashboard built from scratch with Claude Code (Opus) and GitHub Copilot, and through the initial production deployment of **Hive** — an open-source multi-agent orchestration system that realizes Level 6: full autonomy. The system currently operates with 74 CI/CD workflows, 32 nightly test suites, 91% code coverage, and achieves bug-to-fix times under 30 minutes — 24 hours a day. The central finding: the intelligence of an AI-driven development system resides not in the AI model itself, but in the infrastructure of instructions, tests, metrics, and feedback loops that surround it. You cannot skip levels, and at each level, the thing that unlocks the next one is another feedback mechanism. Testing — the volume of test cases, the coverage thresholds, and the reliability of test execution — proved to be the single most important investment in the entire journey. **New in v2:** This revision extends the model from 5 to 6 levels. Level 6 (Fully Autonomous) was introduced 3 weeks after the original publication to distinguish full autonomy — where agents act, merge, and roll back without human gating — from Level 5's semi-automated mode, where agents propose but humans still approve. Hive, the open-source reference implementation for Level 6, was built in the week preceding this revision and is already in production deployment.

---

## 1. Introduction

In December 2025, I started building KubeStellar Console — a Kubernetes multi-cluster management dashboard — from scratch using AI coding agents. Full stack. Go backend, React/TypeScript frontend, Helm charts, CI/CD, the works. I had no team. Just me and Claude Code.

The first two weeks were exhilarating. Code poured out at a pace I'd never experienced. Features that would normally take days appeared in hours. It felt like having a tireless junior developer who typed at the speed of thought.

Then the limitations hit. All at once.

Broken builds. Wrong architectural patterns. Scope creep — the AI modifying files I didn't ask it to touch. And the cascade problem: fix one thing, three others break. I was spending more time reviewing and reverting than I would have spent writing the code myself. The "10x productivity" promise started feeling like a net negative.

This experience — fast initial satisfaction followed by systemic frustration — appears to be universal among developers adopting AI coding tools. The tools are capable, but capability without structure produces chaos. What was missing was not a better AI model. What was missing was a systematic approach to making the AI effective within a real codebase over time.

Several maturity models for AI-assisted development have emerged in 2025-2026 (Section 2.3), but they share a common orientation: defining maturity by how much autonomy AI agents have — how much the human steps back. This paper takes a different approach.

The AI Codebase Maturity Model (ACMM) is a **6-level framework** that defines maturity not by autonomy, but by feedback loop topology — the specific mechanisms through which a codebase measures, adapts to, and governs the behavior of AI agents. Drawing inspiration from CMMI [1], the model argues that each level depends on infrastructure established at the previous level. You cannot skip levels. And the thing that unlocks each transition is not giving AI more freedom — it is adding another feedback mechanism.

The contributions of this paper are:

1. **The ACMM framework** — six levels with defined characteristics, transition triggers, and anti-patterns, applicable regardless of specific AI tooling.
2. **A longitudinal case study** — 4 months of daily AI-assisted development, from zero to a self-sustaining system with quantitative metrics at each level.
3. **Level 6: Full Autonomy** — defined criteria, a reference implementation (Hive, built one week before this revision), and early production evidence demonstrating that multi-agent orchestration with adaptive workload governance can achieve fully autonomous codebase maintenance.
4. **Practitioner guidance** — specific, actionable recommendations for progressing through each level transition.

Have you ever cloned an open source project, used it for a few minutes, and realized it didn't quite solve your challenge? Or waited weeks to find out if a maintainer would even look at your issue? By Level 5 of this model, those experiences become relics. The system I describe processes bug reports to fixes in under 30 minutes, feature requests to implementations in under 60 minutes, 24 hours a day. And at Level 6, the system doesn't just respond to issues — it identifies its own problems, dispatches agents to fix them, merges the results, and notifies the human only when strategic decisions are needed. This is not theoretical. It is running today.

The paper is organized as follows: Section 2 reviews related work in maturity models and AI-assisted development. Section 3 presents the ACMM framework. Section 4 validates Levels 1–5 through the KubeStellar Console case study. **Section 5 introduces Level 6 and the Hive system.** Section 6 discusses key insights and limitations. Section 7 offers implications for practitioners, and Section 8 concludes.

---

## 2. Background and Related Work

### 2.1 Capability Maturity Models

The Capability Maturity Model (CMM) was introduced by the Software Engineering Institute in 1993 [1] to provide organizations with a roadmap for software process improvement. Its successor, CMMI [2], extended the framework across disciplines. The core insight — that process maturity follows predictable, sequential stages, and that each stage depends on capabilities established at the previous stage — has proven durable despite criticism of rigid level definitions [3].

DevOps adopted similar thinking through the DORA metrics framework [4], which identifies four key metrics (deployment frequency, lead time, change failure rate, mean time to restore) that correlate with organizational performance. The DORA model demonstrated that maturity frameworks remain useful when grounded in measurable outcomes rather than prescriptive processes.

## 2.2 AI-Assisted Software Development

The landscape of AI coding tools has expanded rapidly. GitHub Copilot [5], Claude Code [6], Cursor, and autonomous agents like SWE-Agent [7] represent a spectrum from suggestion-based assistance to fully autonomous coding. Empirical studies have shown productivity improvements: Peng et al. [8] found that Copilot users completed tasks 55.8% faster; Ziegler et al. [9] found that suggestion acceptance rate is the primary driver of developers' perceived productivity with AI code completion.

However, these studies focus on individual task completion. They do not address how teams or codebases systematically improve their AI integration over time. A developer who is 55% faster at generating code but spends equivalent time debugging AI-introduced regressions has not achieved a net improvement.

## 2.3 Emerging AI Development Maturity Models

Several frameworks addressing AI-assisted development maturity have appeared in 2025-2026:

**Dan Shapiro's Five Levels** [14] maps progression from Manual (Level 0) to the "Dark Factory" (Level 5), borrowing from the automotive industry's autonomous driving framework. The model focuses on autonomy — how much can AI do without human involvement? — and has been widely adopted in industry discourse [15, 16].

**The AI-SDLC Maturity Model** [17] (ELEKS) defines five stages from Traditional to AI-Autonomous development, with emphasis on enterprise governance and the organizational changes required at each stage.

**The AI Development Maturity Model (AIDMM)** [18] describes five levels from purely human coding to fully autonomous AI-driven codebases, focusing on how the developer's role evolves from writing code to orchestrating agents.

**AI-MM SET** [19] (Gigacore) introduces a three-axis scoring system — Autonomy, Controls, and Governance — arguing that higher autonomy without stronger controls is a risk, not progress.

**Table 0: Comparison of AI Development Maturity Models**

| Model | Defining Axis | Scope | Levels | Validated Case Study | Feedback Loop Focus |
|---|---|---|---|---|---|
| Shapiro / Dark Factory [14] | Autonomy | Individual/team | 6 (0–5) | No (conceptual) | No |
| AI-SDLC [17] | SDLC maturity | Enterprise | 5 | No (consulting) | Governance-focused |
| AIDMM [18] | Developer role evolution | Individual | 5 | No (conceptual) | No |
| AI-MM SET [19] | Autonomy × Controls × Governance | Team/org | 5 | No (framework only) | Controls axis |
| **ACMM (this paper)** | **Feedback loop topology** | **Codebase** | **6** | **Yes (4 months L1–L5 quantitative + 1 week L6 reference implementation)** | **Central thesis** |

These frameworks share a common orientation: they define maturity primarily in terms of how much autonomy AI agents have, or how mature the organization is in adopting AI. They ask: "How much can the AI do without you?"

The ACMM occupies a distinct space: it asks "What must the codebase encode before the next level of AI effectiveness is possible?" The unit of analysis is the codebase and its surrounding infrastructure, not the team or the organization. This makes the model applicable to solo maintainers and large teams alike — the levels are properties of the system, not of the people operating it.

## 2.4 The Gap

The existing models are valuable contributions, but they share a limitation: they are defined by the degree of human absence rather than by the infrastructure that makes AI effective. A team that gives AI full autonomy without measurement is not mature — it is reckless. Autonomy is an output of maturity, not a definition of it.

None of the existing frameworks are defined by feedback loop topology — the specific mechanisms (tests, acceptance metrics, error monitoring, self-tuning configurations) that must exist before the next level becomes possible. None provide a validated case study with quantitative metrics across all levels. None address the concept of the codebase as model — where the code patterns, instruction files, and test suites collectively teach AI agents how to operate, making the repository itself the primary source of AI intelligence rather than the underlying language model. **And none provide a reference implementation for the highest level of autonomy.**

The ACMM fills this gap by defining maturity in terms of what the codebase measures, encodes, and enforces — not in terms of what the human stops doing. With the addition of Level 6 and Hive, the ACMM is the first framework to provide both a complete theoretical model and an open-source reference implementation for fully autonomous AI-driven development, with early production evidence.

---

## 3. The AI Codebase Maturity Model

The ACMM rests on three design principles:

1. **Each level subsumes the previous.** You cannot skip levels. Level 4 requires Level 3's measurement infrastructure, which requires Level 2's consistency, which requires Level 1's basic capability.
2. **Levels are defined by feedback loop topology, not by tool choice.** The same model applies whether you use Claude, Copilot, or any future AI coding tool.
3. **Progression is measured by the degree to which human judgment is encoded in system artifacts** — instruction files, tests, metrics, workflow rules — rather than held in human memory.

### Level 1 — Assisted: Prompt and Review

**Characteristic:** The human initiates every interaction. AI is a sophisticated autocomplete. No persistent context exists between sessions.

**Typical practices:** Ad-hoc prompting, manual review of all AI output, copy-paste from chat interfaces, one-off code suggestions accepted or rejected in isolation.

**Key artifacts:** None beyond the code itself. All knowledge about preferences, patterns, and architectural decisions lives in the developer's head.

**Transition trigger:** The developer notices they keep repeating the same corrections across sessions and wishes the AI "just knew" their preferences.

**Anti-pattern:** "I tried Copilot but it doesn't understand our codebase" — blaming the tool for the absence of encoded context.

### Level 2 — Instructed: Encoded Preferences

**Characteristic:** Preferences, conventions, and architectural decisions are written into files that AI agents consume at the start of every session. Output becomes consistent across sessions and agents.

**Typical practices:** CLAUDE.md files with project rules, .github/copilot-instructions.md with PR and commit conventions, development guides encoding common rejection reasons, structured PR templates with AI-readable checklists.

**Key artifacts:** Instruction files, codified style guides, card/component development guides. In my case, a single card development guide encoded approximately 90% of the reasons I had been rejecting AI-generated PRs.

**Transition trigger:** The developer realizes instructions improve consistency but cannot tell them whether the AI is actually performing well. Gut feeling replaces data. They want measurement.

**Anti-pattern:** Ever-growing instruction files with contradictory rules, no mechanism to determine which instructions are effective.

### Level 3 — Measured: Feedback Becomes Visible

**Characteristic:** The system produces quantitative signals about AI agent performance. Acceptance rates, coverage metrics, error rates, and user feedback are tracked systematically.

**Typical practices:** PR acceptance rate tracking per category, code coverage gating on every PR, nightly test suites (compliance, performance, security, accessibility), GA4 or equivalent error monitoring, NPS surveys for user sentiment, weekly flaky test analysis.

**Key artifacts:** Acceptance rate logs (e.g., `auto-qa-tuning.json`), coverage reports, monitoring dashboards, error classification systems.

**Testing as the breakthrough:** This is where I want to be emphatic. Testing — the sheer volume of test cases, the coverage thresholds, and critically the reliability of test execution — was the single most important investment in the entire journey. At Level 2, I had instructions that made the AI consistent. At Level 3, I had tests that made the AI trustworthy. A flaky test in a human workflow is annoying. A flaky test in an autonomous workflow is dangerous — it erodes the trust that enables every subsequent level. The compounding effect of volume, coverage, and determinism cannot be overstated.

**Transition trigger:** The developer sees patterns in the data and realizes certain responses should be automated — “why am I manually adjusting weights when the data tells me what to do?”

**Anti-pattern:** “Dashboard graveyard” — metrics collected but never acted upon. The human remains the bottleneck for every interpretation and decision.

### Level 4 — Adaptive: Feedback Loops Close Themselves

**Characteristic:** The system acts on its own metrics. Thresholds trigger automated responses. Human oversight shifts from execution to governance.

**Typical practices:** Self-tuning rotation weights based on PR acceptance rates, automated issue triage loops (every 15 minutes across multiple repos), overnight autonomous bug fixes, proactive scanning for user engagement opportunities, contributor leaderboards for coordination, error recovery with exponential backoff.

**Key artifacts:** Self-modifying configuration files (like `auto-qa-tuning.json`, where categories with acceptance rates below 20% are automatically blocked), closed-loop CI/CD pipelines, worktree-based concurrent AI sessions, automated content generation (MARP slides with ElevenLabs narration, documentation sync from PR screenshots).

**Transition trigger:** The developer realizes the system's behavior is now primarily determined by its artifacts — tests, configs, workflows — rather than by real-time human decisions. The code is the policy.

**Anti-pattern:** Autonomous action without sufficient guardrails. The system optimizes for a proxy metric and drifts from user value. This is why the transition from Level 3 is critical — you need measurement before you can safely automate.

### Level 5 — Semi-Automated: The System Proposes, Humans Approve

**Characteristic:** The system detects problems and proposes fixes without human initiation. Instructions encode human judgment. Tests encode trust boundaries. Metrics encode priorities. The community steers the project by opening issues; AI agents implement continuously. But humans still approve — the system proposes, it does not merge autonomously.

**Typical practices:** Community-driven issue-to-implementation pipelines (bug to fix in 30 minutes, feature to implementation in 60 minutes), automated documentation and tutorial generation, multi-agent orchestration across repositories, self-improvement cycles where the system analyzes its own merged PRs and updates its guidance.

**Key artifacts:** The entire codebase is the AI's operating manual. Every test is a trust constraint. Every workflow is a policy. Every metric threshold is a prioritization decision. The instruction files, the code patterns, the acceptance rate history, and the workflow rules form a coherent system that any AI agent can read and operate within — without human presence at the keyboard.

At this level, the accumulated size of the codebase — the tests, the instruction files, the workflow rules, the code patterns — all work in concert not just to resolve issues, but to evaluate whether an issue is genuinely a bug or a user misunderstanding. A concrete example from April 2026: a user filed a bug reporting that a Kubernetes cluster was marked "healthy" while pods were in ImagePullBackOff state. The system responded — before any human reviewed it — explaining that cluster health reflects infrastructure health (node readiness, API reachability), which is architecturally separate from workload health (pod status). The user didn't have a bug. They had a misunderstanding of how Kubernetes clusters work. The codebase knew the difference because the tests, the health-check logic, and the documentation all encoded that design decision [see kubestellar/console#5475].

**Transition trigger:** The developer realizes the system proposes well but still waits for human approval on every merge. The bottleneck is no longer AI quality — it's human availability. The question shifts from "can I trust the output?" to "can I let the system act on what it finds?"

**Anti-pattern:** Alert fatigue: generating proposals no one reviews. An unread suggestion is noise, not maturity. The distinction between L5 and L6 is precisely this: at L5, the system proposes; at L6, it acts.

### Level 6 — Fully Autonomous: The System Runs Itself

**Characteristic:** The system acts on what it finds — generates issues, dispatches agents, merges PRs, rolls back failures. Multiple AI agents operate as a coordinated fleet under adaptive workload governance. Humans set policy and audit after the fact. The human role is strategist: setting direction, defining risk boundaries, and reviewing autonomous behavior.

**Feedback loop:** Multi-loop with external orchestration. A supervisor agent coordinates executor agents. A workload governor adjusts cadence based on real-time backlog. Work is claimed via Beads [21], a distributed work ledger, to prevent conflicts. Push notifications escalate decisions that require human judgment.

**Typical practices:** - Automated issue generation (cron-triggered workflows scanning for TODOs, stale dependencies, failing tests, coverage gaps) - Multi-agent orchestration with role specialization (scanner, reviewer, architect, outreach) - Adaptive workload governance (SURGE/BUSY/QUIET/IDLE modes based on issue/PR backlog) - Merge queue with automated merge (verified PRs merge without manual intervention) - Strategic dashboard (real-time visibility into agent status, token burn rates, governor mode) - Risk assessment configuration (high-risk paths require human review regardless of AI confidence) - Production feedback signals (error monitoring creating regression issues) - Rollback drills (documented procedures for reverting autonomous changes)

**Key artifacts:** - **Agent policy files** — each agent reads a markdown policy file on every firing (no restart needed to change behavior) - **Beads work ledger** [21] — agents claim work with `--actor` to prevent duplicate effort, backed by a versioned database - **Governor configuration** — cadence rules that adapt in real time, sourced from an env file on every tick - **Push notification infrastructure** — ntfy, Slack, Discord for human escalation - **Observability runbook** — how humans debug autonomous behavior

**Anti-pattern:** Mistaking autonomy for abandonment. An autonomous codebase still needs a strategist to set direction, or it drifts toward local optima. Level 6 requires the most sophisticated human judgment of all — but that judgment is applied to strategy and policy, not to individual merges.

**Table 1: ACMM Level Summary**

| Level | Name | Feedback Loop | Key Artifacts | Human Role |
|---|---|---|---|---|
| 1 | Assisted | Open loop | None | Executor |
| 2 | Instructed | One-way (human→AI) | Instruction files | Rule-writer |
| 3 | Measured | Two-way (with human interpretation) | Metrics, test suites | Analyst |
| 4 | Adaptive | Closed loop (automated response) | Self-tuning configs | Governor |
| 5 | Semi-automated | Multi-loop (self-improving) | The codebase itself | Operator |
| 6 | **Fully Autonomous** | **Multi-loop with external orchestration** | **Agent fleet + policy files + Beads ledger** | **Strategist** |

---

## 4. Case Study: KubeStellar Console (Levels 1–5)

### 4.1 System Context

KubeStellar Console is an open-source Kubernetes multi-cluster management dashboard, part of the KubeStellar project under the Cloud Native Computing Foundation (CNCF) Sandbox. Development began in mid-December 2025 with a single maintainer (the author) using Claude Code with Opus for primary development and GitHub Copilot for code review.

The current system comprises a Go backend, a React/TypeScript frontend, Helm charts for deployment, and an extensive CI/CD infrastructure. It is maintained across 4 repositories with multiple concurrent AI coding sessions operating via git worktrees. The full source code, including all workflow files, instruction files, and tuning configurations referenced in this paper, is publicly available [12].

### 4.2 Evolution Through the Levels

**Weeks 1–2 (Level 1 — Assisted): The Honeymoon.** Full-stack development from scratch. Go API handlers, React dashboard, WebSocket connections, Helm packaging — all generated through interactive AI sessions. Output was fast and exciting. But every session started fresh with no memory of decisions made in previous sessions. The same architectural mistakes recurred daily.

**Weeks 3–4 (Level 2 — Instructed): Stopping the Bleeding.** The cascade problem ("fix one thing, three others break") forced the creation of CLAUDE.md — initially just a list of things to stop doing. "Don't modify other people's cards." "Always run build and lint before committing." "Use named constants, not magic numbers." This was followed by copilot-instructions.md for PR conventions and a card development guide encoding the top rejection reasons. The effect was immediate. Consistency across sessions improved. The same mistakes stopped recurring. But I had no way to know how well the system was performing in aggregate.

**Weeks 5–8 (Level 3 — Measured): The Testing Investment.** This was the real turning point. I invested heavily in testing infrastructure: 32 nightly test suites covering compliance, performance, dashboard health, nil safety, accessibility, i18n, and visual regression. Code coverage reached 91% across 12 parallel shards. A weekly flaky test analysis workflow identified unreliable tests and auto-created issues. The acceptance rate tracking system emerged here: `auto-qa-tuning.json`, a file that records how many AI-generated PRs are merged vs. closed for each quality category. This data became the foundation for everything that followed. One specific lesson: a Playwright E2E test for drag-and-drop interactions passed 85% of the time. In a human workflow, that's tolerable. In an autonomous workflow where test results gate PR merges, it meant good fixes were randomly blocked and bad ones randomly approved. I spent three days fixing that single test — a timing issue with animation completion in CI. The lesson extended far beyond drag-and-drop: you cannot have autonomous AI development without deterministic tests.

**Weeks 9–12 (Level 4 — Adaptive): The Loops Close.** With measurement in place, automation followed. The Auto-QA system began running 4 times daily with 8 layers of quality checks. PR acceptance rates per category drove automatic weight adjustments — accessibility PRs at 62% acceptance were boosted (0.93 weight), while operator PRs at 8% acceptance were blocked entirely (0 weight). The system learned what I valued without me manually configuring each preference. A triage loop began scanning 4 repositories every 15 minutes. A PR monitor workflow ran every 60 seconds checking build status. Error recovery workflows implemented exponential backoff for stuck agents. A queue processor limited concurrent AI sessions to 2 to prevent interference. GA4 error monitoring began running hourly, querying production analytics for error spikes and auto-creating GitHub issues. A contributor leaderboard was added that gamifies contributions, provides visibility into our IFOS (Interns for Open Source) program, and feeds back to each contributor which components they are focusing on most and least.

**Weeks 13–16 (Level 5 — Semi-Automated): The System Proposes.** The transition to Level 5 was not a single event but a recognition: the system's behavior was now determined by its artifacts, not by my real-time presence. Issues filed at 2 AM were triaged, assigned, fixed, tested, and ready for my review by 6 AM. The weekly self-improvement cycle analyzed merged PRs and updated the Auto-QA's own guidance — "keep changes under 50 lines for best acceptance rate." Automated tutorial creation was added: MARP slides generated from feature descriptions, narrated by ElevenLabs (voice: Rachel), combined with CDP-captured screenshots. Documentation sync workflows scanned merged PRs, captured UI screenshots, and created docs PRs automatically.

The result — and I do not say this lightly — is what I believe to be the world's first fully integrated and fully automated Kubernetes management and orchestration solution. Not because of any single technology, but because all the pieces are wired together and they don't stop running when I go to bed.

### 4.3 Feedback Loop Inventory

The following table enumerates every feedback loop operating in the KubeStellar Console repository, mapped to the ACMM level where it was introduced.

**Table 2: Complete Feedback Loop Inventory (selected)**

| **Feedback Loop** | **ACMM Level** | **Frequency** |
|---|---|---|
| CLAUDE.md instructions | L2 | Every AI session |
| Copilot instructions | L2 | Every Copilot session |
| Card development guide | L2 | Every card PR |
| PR template checklist | L2 | Every PR |
| Build & lint gate | L3 | Every push |
| Code coverage gate (80%) | L3 | Every PR |
| Full coverage suite (12 shards) | L3 | Hourly on main |
| Weekly flaky test analysis | L3 | Weekly |
| Nightly compliance (10 suites) | L3 | Daily 5am UTC |
| GA4 error monitoring | L3 | Hourly |
| NPS surveys | L3 | After 5th session |
| Auto-QA (8 layers) | L4 | 4x daily |
| Auto-QA self-tuning | L4 | Weekly |
| PR acceptance rate tracking | L4 | Continuous |
| Issue triage loop | L4 | Every 15 minutes |
| Automated tutorial generation | L5 | On feature completion |
| Self-improvement analysis | L5 | Weekly |
| Community issue-to-implementation | L5 | 24/7 |
| **Multi-agent orchestration (Hive)** | **L6** | **Continuous** |
| **Adaptive workload governor** | **L6** | **Every 5 minutes** |
| **Strategic dashboard** | **L6** | **Real-time (SSE)** |

This inventory totals 36+ distinct feedback loops. Each one was built to solve a specific problem. Together, they form the nervous system that enables autonomous operation.

## 4.4 Evidence: Specific Cases

**Case A — Autonomous bug fix, external user (3 hours):** Issue #898: an external user (`namasl`) requested GPU drill-down details on February 12. PR #900 was opened, reviewed, and merged within 3 hours — 624 additions across 7 files. The user did not need to write any code.

**Case B — System identifies user misunderstanding (10 minutes):** Issue #5475: a user reported that a cluster was marked "healthy" while pods were in ImagePullBackOff state. The system responded within 10 minutes, explaining that cluster health reflects infrastructure health (node readiness, API reachability), which is architecturally separate from workload health. The user had a Kubernetes knowledge gap, not a bug.

**Case C — AI failure, PR closed in 8 minutes:** PR #5289: Copilot generated a PR claiming to add "drag and drop for new workloads" but the change was labeled size/XS and did not match the issue requirements. Created and closed within 8 minutes. This is the system working correctly — low-quality output is caught and rejected, feeding back into acceptance rate tracking.

**Case D — Auto-QA catches real quality gap (11 minutes):** Issue #5390: the Auto-QA system detected hardcoded user-facing strings that needed i18n extraction. Filed at 02:03, resolved at 02:14 — 11 minutes from detection to fix. No human initiated this; the system found the problem and fixed it autonomously.

**Case E — Aggregate failure data drives self-correction:** The operator category accumulated 129 closed PRs against only 11 merged (8% acceptance rate). The Auto-QA tuning system automatically set the rotation weight to 0, blocking all future operator-category issues. CI resources were redirected to a11y (62% acceptance, weight 0.93). This is the self-tuning feedback loop in action — the system learned what the maintainer values and stopped wasting cycles on what doesn't work.

## 4.5 Quantitative Summary

**Table 3: System Metrics**

| Metric | Value | Level |
|---|---|---|
| GitHub Actions workflows | 74 (22 AI-specific) | L4 |
| Nightly test suites | 32 | L3 |
| Code coverage | 91% across 12 shards | L3 |
| Distinct feedback loops | 36+ (see Table 2) | L2–L6 |
| Bug-to-fix time | ~30 minutes | L4 |
| Feature-to-implementation time | ~60 minutes | L4 |
| Triage loop frequency | Every 15 minutes, 4 repos | L4 |
| PR monitor frequency | Every 60 seconds | L4 |
| GA4 error monitoring | Hourly | L3 |
| Auto-QA quality checks | 4x daily, 8 layers | L4 |
| Concurrent AI sessions | Multiple via git worktrees | L4 |
| Time from zero to Level 5 | ~82 days | — |
| Time from zero to Level 6 | ~93 days | — |

**Table 4: Throughput Acceleration by ACMM Level**

The most striking pattern in the data is not the absolute numbers but the rate of acceleration. Each maturity level unlocks a step change in throughput — the feedback loops compound.

| Period | ACMM Level | PRs merged/day | Issues closed/day | Commits/day |
|---|---|---|---|---|
| Jan (16 days) | L1–L2 | 16 | 4 | 44 |
| Feb (28 days) | L3 | 29 | 8 | 32 |
| Mar (31 days) | L4 | 40 | 32 | 45 |
| Apr (26 days) | L5–L6 | 81 | 150 | 91 |

PRs merged per day increased **5x** from Level 2 to Level 6. Issues closed per day increased **37x**. The April surge — driven by the introduction of Hive's multi-agent orchestration — is particularly notable: throughput doubled in a single month despite no change in the underlying AI model. The infrastructure, not the intelligence, drove the acceleration.

The commit rate shows a different pattern: relatively stable through Levels 2–4 (~35–45/day), then doubling at Level 5–6. This reflects the shift from "one human directing one agent" to "multiple agents operating in parallel under adaptive governance."

**Table 5: Aggregate Data (Jan 16 – Apr 26, 2026, 100 days)**

| Metric | Value |
|---|---|
| Total commits | 5,435 |
| Total PRs merged | 4,499 |
| Total PRs closed (unmerged) | 713 |
| Overall PR acceptance rate | 86.3% |
| Copilot PRs merged / total | 320 / 859 (37.3%) |
| Total issues opened | 5,282 |
| Total issues closed | 5,273 (99.8%) |
| Issues with ai-fix-requested label | 3,604 |
| Auto-QA issues created and resolved | 654 |
| Average PRs merged per day | 45.0 |
| Contributors | 36 (GitHub) |

---

## 5. Level 6: Full Autonomy and the Hive System

### 5.1 The Transition from Level 5 to Level 6

Three weeks after the original v1 publication, I introduced Level 6 to the ACMM framework. The motivation was not theoretical — it was operational frustration.

Level 5 relied on AI agent CLIs' built-in looping constructs (`/loop`, cron registration, scheduled wake-ups) to keep agents running autonomously. In practice, these mechanisms were **unreliable.** Sessions would interrupt themselves with permission prompts. Loops would silently stop firing. Context windows would fill and the agent would lose its instructions. Rate limits would kill a session, and there was no mechanism to restart it. I would wake up to find that every agent had been idle for hours — not because the work was done, but because the loop had broken. The promise of Level 5's "24/7 operation" was real in theory but fragile in practice.

Hive was born from this frustration. The core insight: **stop depending on AI agent CLIs for reliability, and instead use battle-tested Linux infrastructure — systemd, tmux, cron, bash — for the things that must not fail: scheduling, health monitoring, restart, and coordination.** Let the agents do what they're good at (reading code, reasoning about fixes, writing PRs) and let the operating system do what it's good at (keeping processes alive on a schedule).

This also exposed three structural problems that Level 5 cannot solve even with reliable looping:

1. **Throughput ceiling.** A single agent session processes issues sequentially. When the backlog exceeds ~20 open items, the SLA (30-minute issue-to-fix) becomes unreachable. The bottleneck is not the AI's speed — it's the serial execution model.

2. **Role confusion.** A single agent asked to scan, triage, fix, review, merge, and do outreach switches context constantly. The quality of each task degrades because the agent's attention is diluted. Specialized roles produce better output.

3. **No workload adaptation.** At Level 5, cadences are fixed — every 15 minutes, every hour, regardless of how much work exists. During quiet periods, this wastes tokens. During surges, this falls behind. The system cannot match its intensity to the actual demand.

Level 6 addresses all of these by introducing **multi-agent orchestration with adaptive workload governance**, built on reliable OS-level primitives rather than agent-internal scheduling.

### 5.2 Hive: The Reference Implementation

Hive [20] is an open-source multi-agent orchestration system purpose-built for the Level 6 transition, built in the week preceding this revision. It is available at `github.com/kubestellar/hive` under the Apache 2.0 license.

Hive's design philosophy is a deliberate separation of concerns: **reliable Linux infrastructure handles scheduling, health monitoring, and process management; AI agents handle code reasoning, testing, and PR creation.** Systemd services keep agents alive. Cron-driven governors adjust cadence. Tmux sessions provide isolated execution environments. Bash healthchecks detect stalls and trigger respawns. The agents themselves never need to self-schedule — they read a policy file, do the work, and return to an idle prompt. The operating system decides when to wake them.

This architecture emerged from painful experience with agent CLI looping constructs that would silently fail, prompt for input at 3 AM, or lose context after a rate limit. Hive codifies the patterns that emerged organically during the KubeStellar Console project into a reusable system that any project can adopt.

**One command starts everything:**

```
hive supervisor
```

This installs missing tools, starts all agents, sets the kick cadence, launches the supervisor, and starts the web dashboard. No tmux knowledge needed.

### *Architecture*

Hive runs a fleet of AI agents coordinated by a supervisor, each in its own tmux session:

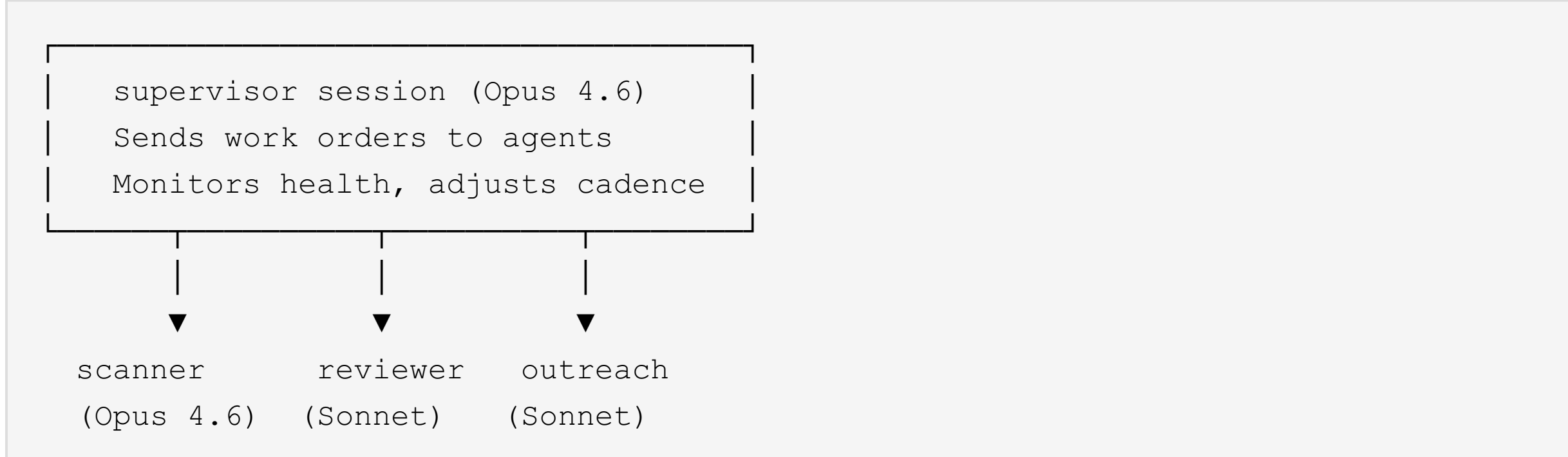

Each agent: - Has its own tmux session and systemd service - Reads a markdown policy file on every firing (no restart needed to change behavior) - Writes to a shared work ledger using `--actor <name>` to claim work - Skips items already claimed by another actor - Notifies the operator via push notifications for decisions requiring human judgment

### *The Adaptive Workload Governor*

The governor measures issue and PR backlog across all managed repositories every 5 minutes and selects a mode:

| Mode | Trigger | Scanner | Reviewer | Architect | Outreach |
|---|---|---|---|---|---|
| SURGE | queue > 20 | 10 min | 10 min | paused | paused |
| BUSY | queue > 10 | 15 min | 15 min | paused | paused |
| QUIET | queue > 2 | 15 min | 30 min | 1 hour | 2 hours |
| IDLE | queue ≤ 2 | 30 min | 1 hour | 30 min | 30 min |

Architect and outreach are **opportunistic** — they fill idle cycles and pause entirely under load. The supervisor runs every 5 minutes regardless of mode. Cadences are tunable in a configuration file — no restart needed.

This is the key insight: **the system matches its intensity to the actual workload.** During surges, all resources focus on clearing the backlog. During quiet periods, the system shifts to improvement work (architecture, outreach). The governor makes this decision autonomously — the human sets the thresholds, not the individual kicks.

### *Two Scheduling Models*

Hive supports two fundamentally different ways to drive an agent:

**Model A — Self-scheduling:** The agent registers its own cron job and fires on a fixed cadence indefinitely. The supervisor's only job is to keep the session alive via systemd. Low operator involvement. Best for single-agent setups or batch jobs — but vulnerable to the same loop-reliability problems that motivated Hive's creation.

**Model B — EXECUTOR MODE (supervisor-driven):** The agent starts, reads its policy, then **waits at the prompt** for the supervisor to send work orders via `tmux send-keys`. No cron, no self-scheduling. A systemd timer fires the governor, which decides when to kick each agent and what work to assign. The agent never needs to maintain its own loop — it does the work and returns to idle. This is the key architectural insight: the scheduling reliability problem is moved entirely out of the AI agent and into the OS.

The production deployment uses Model B for all agents. This eliminated the loop-reliability failures that plagued Level 5 — agents no longer silently stop, prompt for input, or lose context. If an agent crashes, systemd restarts it. If it stalls, the healthcheck kills and respawns it. The agent's only job is to be good at code.

***CLI-Backend Agnostic***

Hive does not require a specific AI tool. It supports:

| Backend | Type | Description |
| --- | --- | --- |
| `claude` | Native CLI | Anthropic's CLI — runs Claude models directly |
| `gemini` | Native CLI | Google's CLI — runs Gemini models directly |
| `copilot` | Aggregate | GitHub Copilot — routes to Claude, GPT, Gemini |
| `goose` | Aggregate | Block's Goose — routes to any model via config |

Backends can be swapped per agent at runtime with `hive switch <agent> <backend>`. Optional local model support (Ollama + litellm proxy) enables zero-API-cost operation.

***The Strategic Dashboard***

`hive dashboard` launches a real-time web dashboard on port 3001:

- **Live updates via SSE** — agent states, governor mode, repo counts refresh every 5 seconds
- **Sparkline history** — per-agent busy time and restart count with rolling 24-hour window
- **Intensity gauge** — speedometer comparing recent vs. trailing token burn rates

- **Kick buttons** — one-click immediate kick for any agent
- **Backend switching** — swap agent CLI from the UI with automatic pinning
- **Coverage tracking** — test coverage progress toward configured target
- **macOS Übersicht widget** — desktop widget showing agent status at a glance

The dashboard is the L6 observability artifact: it gives the strategist visibility into autonomous behavior without requiring them to watch tmux sessions.

***Work Coordination: Beads***

Multi-agent coordination requires a shared state mechanism that agents can read and write without conflicting. Hive uses **Beads** [21], an open-source distributed issue tracker by Steve Yegge, backed by Dolt (a version-controlled SQL database). Beads provides exactly what L6 needs: structured work items with actor ownership, dependency chains, and a full audit trail — all stored in a Git-like versioned database that agents can query and update from the command line.

Each agent: - Claims work items with `bd add --actor <name>` before starting - Queries for unclaimed work: `bd list --status=open` - Skips items already claimed by another actor: `bd list --actor=<other> --status=in_progress` - Records completion status and notes for audit: `bd update <id> --status done --notes "..."`

This prevents the most dangerous L6 failure mode: two agents fixing the same issue simultaneously, creating conflicting PRs. The versioned backing store also provides a complete audit trail of who worked on what and when — essential for post-hoc review of autonomous behavior.

Beads also solves a subtler problem: **agent memory continuity.** AI agent CLIs lose context in multiple ways — conversation compaction discards older messages, session restarts clear the context window, switching CLI backends (e.g., from Claude to Copilot) starts a fresh session with no memory of prior work, and rate limits can kill a session mid-task. Without an external state store, the incoming agent has no idea what the outgoing agent was working on, what it already tried, or what it decided. Beads provides that continuity. When an agent starts — regardless of which CLI backend it runs on, whether it was restarted, or whether its predecessor was compacted — it reads the Beads ledger and immediately knows: what work exists, who claimed it, what's in progress, and what's done. The agent's "memory" lives in the ledger, not in the conversation history. This makes the system resilient to every failure mode that plagues long-running agent sessions.

***Health Monitoring and Self-Healing***

Hive implements four layers of resilience:

1. **Supervisor service** — polls every 10 seconds for agent process crashes, tmux session death, TUI-ready detection

2. **Renew timer** — every 6 days, kills and restarts agent sessions (Claude Code's `/loop` cron expires at 7 days)
3. **Healthcheck timer** — every 20 minutes, checks heartbeat file freshness. Stale agents are killed and respawned
4. **Push notifications** — ntfy, Slack, Discord alerts on stalls, recoveries, and escalations

After `MAX_RESPAWNS` failed attempts, the system stops auto-respawning and pages the human with "manual intervention needed."

## 5.3 Level 6 Criteria

The following criteria define what a codebase needs to achieve Level 6:

| Criterion | Description | Rationale |
| --- | --- | --- |
| Automated issue generation | Cron-triggered workflow that scans for TODOs, stale deps, failing tests, coverage gaps | The codebase proposes its own next task |
| Multi-agent orchestration | Dispatcher coordinating multiple AI agents on parallel work | Single agents become units of a larger system |
| Merge queue / auto-merge | Verified PRs merge without human intervention | Humans gate the queue config, not individual merges |
| Strategic dashboard | Shows autonomous activity — sessions, fixes, merge pipeline | Visibility into autonomous behavior |
| Risk assessment config | Prevents autonomous changes to high-risk areas | Blast radius awareness |
| Production feedback signal | Production observations feed back into development | Closes deployment-development gap |
| Observability runbook | Guide for debugging autonomous behavior | Humans need to understand and override |
| Rollback drill | Documented procedure for reverting autonomous changes | Autonomy without a kill switch is recklessness |

## 5.4 Early Production Results (First Week of Hive)

Hive was deployed on the KubeStellar project in its first week, managing 6 repositories with a target SLA of <30 minutes from issue filed to PR merged. These are early results — the system is days old, not months — but they demonstrate that the Level 6 infrastructure works in production.

**Table 6: Hive Early Production Metrics**

| Metric | Value |
|---|---|
| Issues closed autonomously | 36+ |
| PRs merged autonomously | 20+ |
| Median fix time | ~20 minutes |
| Repositories under management | 6 |
| Agent roles | 5 (scanner, reviewer, architect, outreach, supervisor) |
| Production incidents from autonomous fixes | 0 |
| Categories fixed | test failures, CI config, accessibility, z-index bugs, broken deep-links, stale formulas, doc inaccuracies, nightly workflow failures |

**Zero production incidents caused by automated fixes.** Even in the first week, this validates that the L1–L5 infrastructure — the tests, the coverage gates, the acceptance rate tracking — provides sufficient guardrails for fully autonomous operation. Longer-term data will follow as the deployment matures.

### 5.5 Early Lessons from Level 6

**Never trust the agent to keep itself alive.** The single most important lesson from the Level 5 → Level 6 transition: AI agent CLIs are reasoning engines, not process managers. Their built-in looping and scheduling constructs (`/loop`, cron registration, scheduled wake-ups) are convenient for interactive use but unreliable for autonomous operation. Sessions interrupt with permission prompts, loops silently stop firing, context windows fill and instructions are lost, rate limits kill sessions with no recovery mechanism. Hive's architecture — systemd for process management, cron for scheduling, bash for healthchecks, tmux for isolation — eliminated an entire class of failures that no amount of prompt engineering could fix.

**Separation of scanning and fixing pays off.** The scanner runs even when AI sessions are down (restarting, usage-limited, rate-limited). State is never lost. When the AI comes back, it reads the database and knows exactly what happened while it was gone. This resilience proved critical during Claude Code session restarts, model rate limits, and network issues.

**Policy files > embedded instructions.** Each agent reads its policy file fresh on every firing. This means behavior changes take effect immediately — no session restart, no redeployment. The policy is the single source of truth for agent behavior.

**Backoff prevents thrashing.** The `fix_attempts` counter in the state database is critical. Without it, the fix-loop retries the same unfixable issue every 15 minutes forever. With it: 3 failed attempts → status='skip', operator gets notified, issue stays tracked but stops consuming agent time.

**The governor is the key differentiator.** Fixed cadences (Level 5) waste resources during quiet periods and fall behind during surges. The adaptive governor matches system intensity to actual demand. This is the difference between "the system runs on a schedule" and "the system responds to what's happening."

**External state survives everything the agent doesn't.** Beads proved essential not just for coordination but for continuity. Agent sessions die — from compaction, restarts, CLI switches, rate limits, crashes. Each death erases the agent's conversational memory. But the Beads ledger persists on disk, versioned by Dolt, readable by any agent on any CLI backend. An agent that restarts after a crash reads `bd list` and picks up exactly where its predecessor left off. This decoupling of work state from agent memory is what makes L6 robust enough to run unsupervised. Without it, every session restart is a cold start — the agent has to re-scan GitHub, re-discover what's in progress, and risk duplicating work another agent already claimed.

---

## 6. Discussion

### 6.1 The Intelligence Is in the System, Not the Model

The AI model — whether Claude Opus, GPT-4, or any successor — is a commodity component within this architecture. The differentiation lies entirely in the surrounding infrastructure: the instruction files, the test suites, the feedback loops, the monitoring workflows, the tuning configurations. Switching from one AI model to another would require modest effort. Rebuilding the surrounding system would require months.

This has a practical implication: organizations investing in "AI strategy" should invest primarily in the infrastructure of intelligence — the tests, the metrics, the feedback loops — not in selecting the perfect model.

### 6.2 Ask Questions, Not Commands

One operational insight proved disproportionately valuable: prompting with "why didn't you catch this?" rather than "fix this bug." The former produces root cause analysis and systemic improvements — a new test, a new rule in the instruction files, a pattern that prevents the entire class of error. The latter produces a patch. Over time, questioning compounds into a self-improving system; commanding produces a sequence of isolated fixes.

This is the operational manifestation of the Level 1→2 transition: questions produce instructions as a side effect.

### 6.3 Refactoring as Imperative

In traditional development, refactoring competes with feature work for engineering time. In AI-augmented codebases, this calculus inverts. AI agents perform measurably better on clean, well-structured code. Technical debt directly degrades AI output quality, manifesting as lower acceptance rates, more cascading failures, and longer review cycles. Refactoring is not a tax — it is an investment that pays returns on every subsequent AI session.

### 6.4 Telemetry as the Nervous System

The telemetry layer gives immediate, continuous feedback on engagement, reach, and — most importantly — failures and errors. GA4 tracks custom events (`ksc_error` with category, page, and detail dimensions), NPS surveys capture user sentiment, the analytics dashboard visualizes funnels from landing to agent connection to mission completion. When something breaks in production, the hourly error monitor catches it, creates an issue, and an agent is already working on a fix. This telemetry isn't a nice-to-have bolt-on — it is the sensory input that makes the entire feedback loop architecture functional. Without it, the system would be autonomous but blind.

### 6.5 Community-Steered Open Source

At Level 5, something unexpected emerged: the project became what I consider the purest form of community-steered open source. Users don't need to write code. They don't need to understand Go or TypeScript. They open an issue describing what they want, and the system builds it. Bug reports are resolved in 30 minutes. Feature requests are implemented in an hour. Twenty-four hours a day, seven days a week.

The community doesn't just influence the roadmap — they are the roadmap. Every issue is a direction. Every piece of feedback is a steering input. This may represent a new model for open source maintenance that addresses the chronic maintainer burnout problem: community-steered, AI-implemented, human-governed.

### 6.6 The Level 5 → Level 6 Transition

The transition from Level 5 to Level 6 is qualitatively different from all previous transitions. Levels 1 through 5 are about making a single AI agent progressively more effective within a codebase. Level 6 is about **orchestrating multiple agents as a system.** The jump is analogous to the difference between optimizing a single server and building a distributed system — the problems change in kind, not just in degree.

The new problems that emerge at Level 6: - **Scheduling reliability:** Agent CLI looping constructs fail silently — systemd and cron do not - **Work coordination:** Preventing duplicate effort across agents - **Priority allocation:** Deciding which agent works on what - **Workload adaptation:** Matching system intensity to actual demand - **Fleet health:** Keeping multiple agents alive, on-task, and on the right backend - **Escalation protocols:** Knowing when a decision exceeds agent authority

These are infrastructure problems, not intelligence problems — which is consistent with the ACMM's central thesis. The solution is not a smarter model. The solution is better orchestration infrastructure. Specifically, the solution is recognizing that AI agents and operating systems are good at different things, and composing them accordingly: Linux for reliability, agents for reasoning.

### 6.7 Limitations

This paper reports on a single case study with a single maintainer and a specific technology stack. KubeStellar Console is primarily a dashboard (UI/API integration), not safety-critical or algorithmically complex software — generalizability to other domains is an open question. The solo maintainer context may not directly map to large teams, though the model's levels are defined by system properties rather than team properties. There is inherent survivorship bias: I report on a system that reached Level 6 and do not observe systems that failed at intermediate levels. The maturity model has not been independently validated across multiple organizations.

The Level 6 deployment (Hive) has been in production for approximately one week at the time of writing. The early results are promising — 36+ issues closed autonomously with zero production incidents — but one week is not sufficient to draw conclusions about long-term autonomous behavior. Longer-term data on edge cases, drift, maintenance burden, and failure modes at scale remains to be collected and will be the subject of future updates.

---

## 7. Implications for Practice

**For individual developers:** Start with Level 2. Writing a CLAUDE.md file takes 30 minutes and immediately improves AI output consistency. Then invest in test coverage — not just for correctness, but as the trust mechanism that enables every subsequent level. Don't overthink it. Get it working, normalize things later.

**For engineering leaders:** Assess your current level honestly. Most organizations are at Level 1–2. The transition to Level 3 requires investment in measurement infrastructure — acceptance rate tracking, coverage gating, error monitoring. Budget for the "infrastructure of intelligence," not just AI tool licenses.

**For the open source community:** Level 5 suggests a new maintenance model: community-steered, AI-implemented, human-governed. Standard instruction file formats (CLAUDE.md, copilot-instructions.md) could become as important as README.md and CONTRIBUTING.md. Level 6, with Hive as a reference implementation, makes this model available to any project willing to invest in the prerequisite infrastructure.

**For researchers:** The model needs multi-case validation. I invite researchers to apply ACMM to other projects and report findings. Open questions include: Does the model apply to safety-critical systems? How do large teams progress through levels? What are the economic inflection points for each transition? How does Level 6 behavior change over periods longer than 30 days?

**For anyone starting from scratch:** Hive exists so you don't have to build the Level 6 infrastructure yourself. `curl | bash`, edit one config file, run `hive supervisor`. The hard problem is Levels 1–5 — the tests, the instructions, the feedback loops. That work is irreplaceable and specific to your codebase. Level 6 orchestration is generic infrastructure, and it should be shared. The key architectural lesson: don't fight your agent CLI's flaky looping constructs — use systemd, cron, and tmux for reliability, and let the agents focus on what they're good at: reasoning about code.

---

## 8. Conclusion

This paper presented the AI Codebase Maturity Model (ACMM), a **6-level framework** for understanding how codebases evolve from basic AI-assisted coding to fully autonomous systems. The model was validated through a 4-month practitioner experience report building KubeStellar Console from scratch with AI coding agents for Levels 1–5, and through the initial production deployment of Hive — a multi-agent orchestration system built in one week that realizes Level 6.

The central finding is that systematic investment in the infrastructure surrounding AI tools — instruction files, test suites, metrics, feedback loops — yields compounding returns. Each maturity level depends on feedback mechanisms established at the previous level. You cannot skip levels. And at every level, the thing that unlocks the next one is another feedback mechanism.

Testing proved to be the single most important investment: the volume of test cases, the coverage thresholds, and the reliability of test execution together form the trust foundation that makes autonomous AI development possible.

**New in v2:** Level 6 proves that the model extends beyond semi-automated to fully autonomous. When you compose multiple AI agents under adaptive workload governance, with explicit work coordination, risk-aware merge policies, and strategic dashboards for human oversight — the system runs itself. Humans set direction. The codebase executes.

Most developers and organizations using AI for coding are at Level 1 or 2. That is where everyone starts. The point is not to race to Level 6. The point is to know what the next level looks like and what feedback mechanism unlocks it.

The code remembers what I forget. The tests catch what I miss. The feedback loops keep the system honest. Systemd keeps the agents alive. The agents coordinate what I cannot oversee. But the values, the priorities, the decision about what to build and what to skip — that's still me.

---